\documentclass[a4paper]{jpconf}
\usepackage{graphicx}
\begin{document}
\title{Numerical characterization of the ARAPUCA: a new approach for LAr scintillation light detection}

\author{F Marinho$^1$, L Paulucci$^2$, A. A. Machado$^2$ and E  Segreto$^3$}
\address{$^1$ Universidade Federal de S\~ao Carlos, Rodovia Anhanguera, km 174, 13604-900, Araras, SP, Brazil}
\address{$^2$ Universidade Federal do ABC, Av. dos Estados, 5001, 09210-170, Santo Andr\'e, SP, Brazil}
\address{$^3$ Instituto de Física Gleb Wataghin, Universidade Estadual de Campinas, Rua S\'ergio Buarque de Holanda, 777, 13083-859, Campinas, SP, Brazil}

\ead{fmarinho@ufscar.br}

\begin{abstract}
The ARAPUCA concept has been proposed as a simple and neat solution for increasing the effective collection area of SiPMs through the shifting and trapping of scintillation light in noble liquids, thus with great potential for improving timing and calorimetry resolution in neutrino and dark matter search experiments using time projection chambers. It is expected to achieve a single photon detection efficiency larger than 1\%. The initial design consists of a box made of highly reflective internal surface material and with an acceptance window for photons composed of two shifters and a dichroic filter. The first shifter converts liquid argon scintillation VUV light to a photon of wavelength smaller than the dichroic cutoff, so the surface is highly transparent to it. When passing through the dichroic filter, it reaches the second shifter which allows the photon to be shifted to the visible region and be detected by the SiPM nested inside it. When it enters the box, the photon will likely reflect a few times, including on the dichroic filter surface, before being detected. We present a full numerical description of the device using a Monte Carlo framework, including characterization of the acceptance window, models of reflection of different materials, and sensor quantum efficiency, that can now be used to further improve the detection efficiency by comparing different geometries, positions of SiPM and materials. Estimates of simulated efficiencies, number of reflections and acquisition time are presented and compared to analytical calculations. Those are very promising results, giving a total efficiency for the detection of scintillation light in liquid argon of 1.7$\rm \pm$0.3\%. Comparison of the estimated total efficiency with a preliminary result from an experimental test with an ARAPUCA prototype made in Brazil is also presented. 
\end{abstract}

\section{Introduction}
Investigations at the frontier of particle physics involve measurements of low frequency events, such as interacting neutrinos, possible proton decay, and detection of dark matter. New technologies are in high demand to improve the chance of detecting such events. Scintillation light detectors are important component systems for the next generation of LArTPCs experiments which can provide additional information regarding calorimetry and timing thus enhancing event reconstruction performance. To explore the full potential of these systems one must consider the best options for detector technologies under development and their impact on quality of physics measurements. 

One recent suggestion for photon detection at the VUV range is the use of ARAPUCA devices \cite{Machado2016} for trapping photons and increasing the detection effective area of silicon photomultipliers (SiPMs). The concept consists of a box cavity with highly reflective internal walls inside which photons can reflect back and forth until reaching its photomultipliers. This box is equipped with an acceptance window formed by two layers of wavelength shifters and a dichroic filter which allows photons to pass efficiently inside the cavity. It also presents high internal reflectance to the accepted photons, therefore acting as a trap. For use in a liquid argon (LAr) environment, important for dark matter and neutrino searches, those layers were optimized as follows. The scintillation light emitted by argon atoms, in the vacuum ultra-violet spectral region (VUV) with a wavelength around 127 nm, are converted to the detector's range (visible). The first shifter in the acceptance window, made of p-terphenyl (PTP) \cite{DeVol1993}, converts the liquid argon scintillation light to a photon of wavelength that peaks around 350 nm, smaller than the dichroic cutoff ($\sim$ 400 nm). When passing through the dichroic filter, it reaches the second shifter, made of Tetraphenyl-butadiene (TPB) which shifts to the visible region (peaking at $\sim$ 430 nm) \cite{Francini2013}, ensuring the dichroic filter surface to now become highly reflective. The photon can reflect a few times on the internal surface before being detected by the SiPM.

We present a Monte Carlo based approach to characterize the ARAPUCA light scintillation detector device as a tool for research and development of this technology. 

\section{Numerical simulation}

\begin{figure}
\centering
\includegraphics[width=0.3\textwidth]{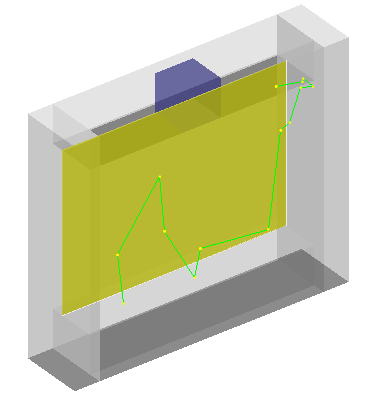}
\includegraphics[width=0.65\textwidth]{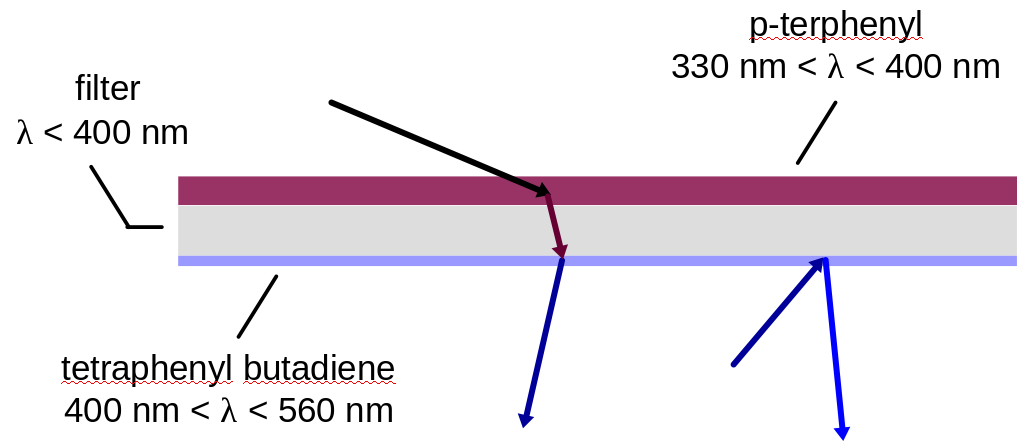}
\caption{ARAPUCA device geometry (left): acceptance window (yellow), SiPM sensor (blue) and reflective cavity (gray). Acceptance window concept (right): PTP (dark magenta), dichroic filter (gray), TPB (blue). }
\label{ara1}
\end{figure}

The computational description of the ARAPUCA was developed in order to include its main components and their correspondent optical functions correctly. For that, the Geant4 simulation framework \cite{Agostinelli2003, Allison2016} provides a suitable set of functionalities that allows the propagation of optical photons through various materials and interfaces considering different models for light reflection, refraction, absorption, emission, etc. Figure \ref{ara1} shows a complete 3D model of a detector indicating its acceptance window, reflective cavity and a silicon photomultiplier (SiPM). The green line indicates a photon optical path and the yellow dots its reflection points on the cavity and filter internal reflective walls.

\begin{figure}
\centering
\includegraphics[width=0.49\textwidth]{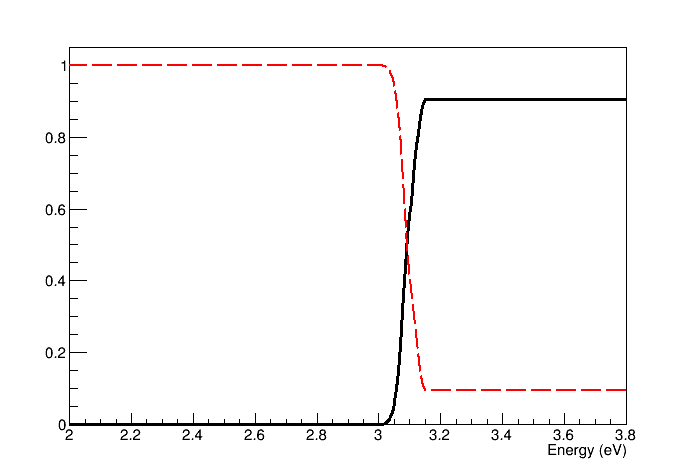}
\includegraphics[width=0.49\textwidth]{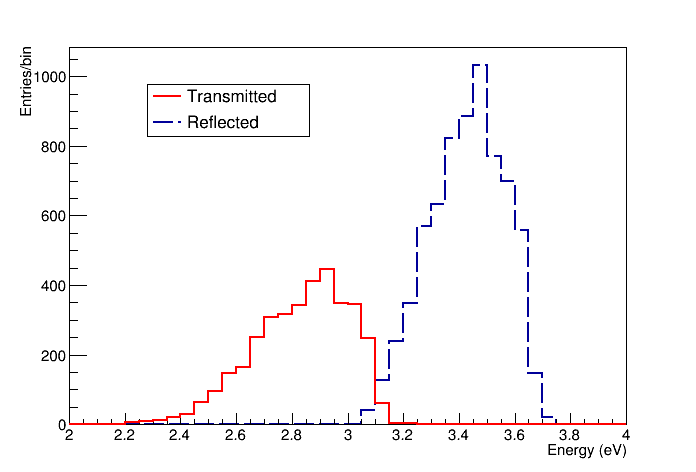}
\caption{Dichroic filter (left): Transmission (solid black) and reflection (dashed red) \cite{filter}. Window emitted photon spectra (right).}
\label{ara2}
\end{figure}

The acceptance window consists of a dichroic filter with one layer of $\rm\sim0.3 mg/cm^2$  PTP deposited on top and one layer of $\rm\sim7\times 10^{-2} mg/cm^2$ TPB  deposited on the bottom of the filter as illustrated in figure \ref{ara1}. The widths of the PTP and TPB are chosen to ensure optimal wavelength shifting  \cite{DeVol1993}. The incoming VUV photons are absorbed by the PTP material which in turn emits photons  isotropically in the 330-400 nm range. As a consequence, half of these emitted photons pass through the dichroic material, reach the TPB and are absorbed. The TPB emits photons in the 400-560 nm range, which therefore, are trapped within the cavity. Figure \ref{ara2} shows the average transmittance and reflectance of the dichroic filter at $45^o$ incidence angle (left) and the spectra for the accepted and reflected light (right) both as a function of the photons' energy.

For the internal walls there are two main possibilities for modelling the reflection types occurring due to the materials envisioned for the device. The box can be made of $\rm Teflon^{TM}$ PTFE which offers high reflection ($ R > 95\%$) with Lambertian angular distribution, however the minimum thickness of the walls is limited due to the need to avoid transmission. An alternative is to use some adequate material for mechanical support and coating the internal walls with $\rm Vikuiti^{TM}$ 3M which is a specular reflective film ($ R \sim 98\%$) \cite{vikuiti}. The  trajectory of the photons inside the cavity is fully described in the simulation which allows to calculate the time interval between photon acceptance and detection at the SiPM surface. This is an important parameter for the detector timing evaluation. 

\begin{figure}
\centering
\includegraphics[width=0.55\textwidth]{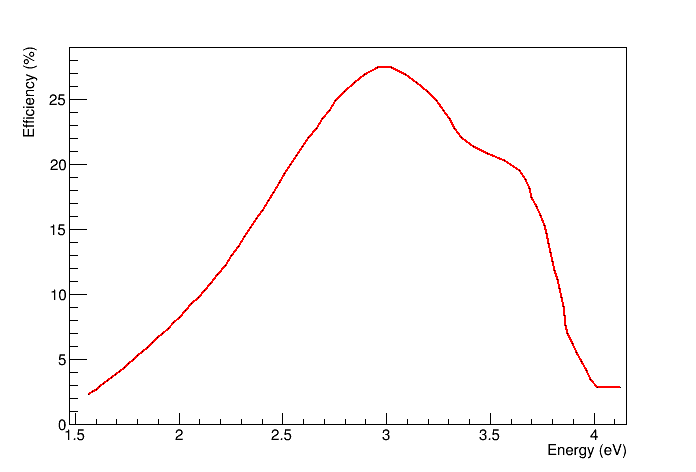}
\caption{SiPM photon detection efficiency as function of incoming photon energy.}
\label{ara3}
\end{figure}

Photon detection efficiency dependence with photon energy must be taken into account for the SiPM sensor. In particular, we used Sensl C-Series models with efficiency shown as function of energy in figure \ref{ara3} in our model \cite{sensl}.   

\section{Results}

A detailed description of the photons trajectories within the different materials of the device geometry allows to evaluate all the relevant parameters for its characterization and overall operation. The following sections describe results obtained with the complete simulation of the ARAPUCA or its components. Comparison with analytic calculations or measurements is provided when those are accessible.

\subsection{Efficiency}

The total efficiency of the detector can be defined as:

\begin{equation}
\epsilon_{total} = \frac{N_{detected}}{N_{total}},
\end{equation}
where $N_{detected}$ is the number of observed photons from the SiPM signal and $N_{total}$ is the number of incoming photons which arrive at the acceptance window from outside.

However, it is important to characterize each component in terms of its own efficiency. Hence, one can factorize the total efficiency as:

\begin{equation}\label{eq}
\epsilon_{total} \approx \epsilon_{acceptance} \times \epsilon_{collection} \times \epsilon_{SiPM},
\end{equation}
where $\epsilon_{acceptance}$ is the window efficiency on converting incoming VUV photons to visible light inside the cavity and $\epsilon_{collection}$ gives the fraction of accepted photons which reach the SiPM (i.e. collected and not absorbed in the cavity walls). The SiPM efficiency is given by:

\begin{equation}
\epsilon_{SiPM}=\frac{1}{N_{collected}}\int_{E_0}^{E_1}\epsilon_{SiPM}(E)\frac{dN_{collected}}{dE}dE,
\end{equation}
where $N_{collected}$ is the number of photons which arrive at the surface of the SiPM and $\epsilon_{SiPM}(E)$ is the sensor photon detection efficiency as a function of the photon energy.

Note the factorization in equation \ref{eq} is possible because $\epsilon_{acceptance}$ is constant around the VUV incoming photons energy. The $\epsilon_{collection}$ is also assumed constant as it should only depend on $R$, the active coverage $f$ and geometry of the reflective cavity. In both aforementioned materials $R$ is above $\rm 95\%$ and approximately constant across the whole wavelength range inside the cavity. The other two parameters are independent. 

Table \ref{table1} shows the value obtained for these efficiencies calculated for a device with $\rm 36~mm \times 25~mm \times 6~mm$ internal dimensions and $ R = 95\%$. An experimental test realized with a prototype following the same configuration has provided a value $\epsilon_{total}^{exp} \sim 1.8 \%$ which is in reasonable agreement with the Monte Carlo estimate.

\begin{table}[h]
\centering
\caption{Monte Carlo estimated ARAPUCA efficiencies.}
\label{table1}
\begin{tabular}{llll}\hline
  $\epsilon_{acceptance}$ & $\epsilon_{collection}$ &  $\epsilon_{SiPM}$ & $\epsilon_{total}$ \\ \hline
 $34.5\pm 6.0 \%$ & $19.7\pm0.4 \%$  & $25.1\pm 0.3 \%$ & $1.7\pm 0.3 \%$ \\ \hline
\end{tabular}
\end{table}

\begin{figure}
\centering
\includegraphics[width=0.9\textwidth]{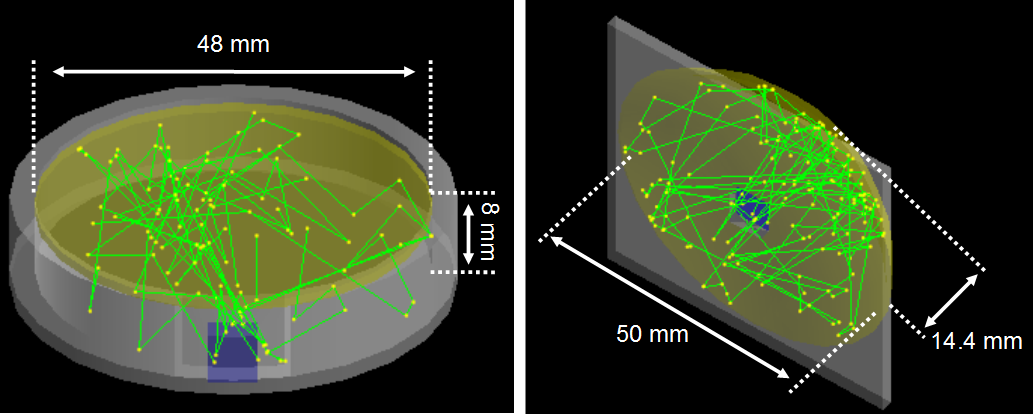}
\caption{Focusing geometries.}
\label{ara4}
\end{figure}

Studies with different geometries for the cavity were also performed such that partial focusing of the light could be used for improving the efficiency. Figure \ref{ara4} shows two possible alternatives which provide $\epsilon_{collection}$ estimates above analytic calculations given by \cite{Machado2016}:
\begin{equation}
\epsilon_{collection}^{analytic}=\frac{f}{1-R(1-f)}.
\end{equation}
Note that it is not possible to obtain efficiencies higher than the analytic estimates with the proposed non-focusing box design from figure \ref{ara1}. The cylindrical cavity with a disk shaped acceptance window provides $\epsilon_{collection} = 14.9\pm0.4 \%$ while the analytic estimate for an equivalent setup (same $R$ and $f$) gives $\rm 13.0\%$. The cavity with a spherical dome shaped window gives $\epsilon_{collection} = 16.0\pm0.4 \%$ and the analytic estimate is $\rm 13.7\%$. These calculations indicate that there is room for improvement of up to 15\% on the overall efficiency with just geometry optimization of the original ARAPUCA proposal.

\subsection{Timing}

The evaluation of the time characteristics of the device is another important aspect that needs to be understood if considering it to be used for timing purposes in a LArTPC experiment. There are three major delay sources that determine the acquisition time of the device:
\begin{equation}
t_{acquisition} = t_{PTP}+t_{TPB}+t_{collection},
\end{equation} 
where $t_{PTP}$ and $t_{TPB}$ are the emission decay time of the wavelength shifters and $t_{collection}$ is the time it takes for the photon accepted in the cavity to arrive on the SiPM after successive reflections. 

\begin{figure}[h]
\centering
\includegraphics[width=0.49\textwidth]{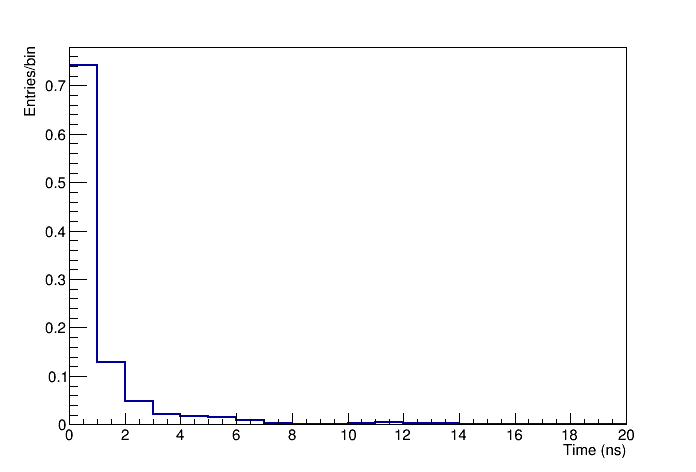}
\includegraphics[width=0.49\textwidth]{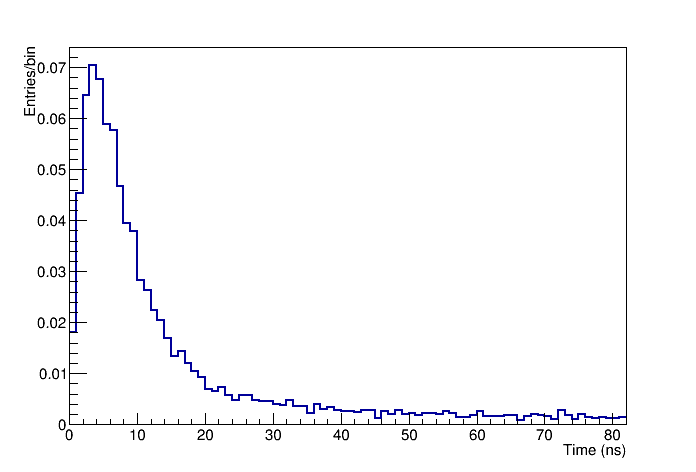}
\caption{Expected distribution for $t_{collection}$ (left) and $t_{acquisition}$ (right).}
\label{ara5}
\end{figure}

All these times are random quantities in a event-by-event basis. Figure \ref{ara5} on the left shows an example of the  $t_{collection}$ distribution obtained from the Monte Carlo simulation. As expected the distribution has a sharp decay as the overall distance traveled is very small inside the cavity. One can evaluate a reasonable distribution for $t_{aquisition}$ assuming a mixture of exponential distributions for each of the emission times and their respective measured values \cite{tpbemi}. Figure \ref{ara5} on the right shows the obtained distribution with a most probable acquisition time of $\rm \sim 4~ ns$. The long tail results mainly from the TPB time response function to the VUV 127 nm photons. In a LArTPC such fast acquisition may allow to determine with good resolution the time $t_0$ for the interacting primary particle in a physics event or to help as trigger device to distinguish beam events from other sources in a surface installed experiment.

\section{Conclusion}

An ARAPUCA full simulation was implemented and is currently employed as a R\&D design tool for this detector technology. The total efficiency obtained from a prototype simulation and its experimental value were in good agreement. Comparison between the analytic estimates and simulation for the collection efficiency, number of internal reflections and timing as function of the $f$ and $R$ parameters follow the same behavior. It is a flexible software application which allows not only parameter optimization but geometrical and physical modifications with ease. In particular one can evaluate the performance of each component according to its features such as wavelength shifters types, width, spectra emission dependence on temperature, dichroic filter transmission and reflection as function of the angle, etc. One can also choose the main surface models according to the type of cavity walls and adjust the response of the SiPM used according to the operating parameters. In addition a set of analysis algorithms were also developed to evaluate all the relevant figures of merit of interest.

\ack
The authors would like to thank Funda\c c\~ao de Amparo \`a Pesquisa do Estado de S\~ao Paulo (FAPESP) for financial support under grants $\rm n^o$ 2016/01106-5 and $\rm n^o$ 2017/13942-5.


\section*{References}
\begin{thebibliography}{99}
\bibitem{Machado2016}
      Machado A A and Segreto E 2016 {\it JINST} ARAPUCA a new device for liquid argon scintillation light
detection {\bf 11} C02004

\bibitem{DeVol1993}
      DeVol T A, Wehe D K, Knol G F 1993 {\it Nucl. Instr. Meth. Phys. Res.} A Evaluation of p-terphenyl and 2,2" dimethyl-p-terphenyl as wavelength shifters for barium fluoride {\bf 327} 354

\bibitem{Francini2013}
      Francini R {\it et al} 2013 {\it JINST}  VUV-Vis optical characterization of Tetraphenyl-butadiene films on glass and specular reflector substrates from room to liquid Argon temperature {\bf 8} P09006

\bibitem{Agostinelli2003}
     Agostinelli S {\it et al} 2003 {\it Nucl. Instr. Meth. Phys. Res.} A Geant4 - a simulation toolkit {\bf 506} 250
    
\bibitem{Allison2016}
      Allison J {\it et al} 2016 {\it Nucl. Instr. Meth. Phys. Res.} A Recent developments in GEANT4 {\bf 835} 186

\bibitem{filter}
       Dichroic filter, edmundoptics.com.

\bibitem{vikuiti}
      $\rm Vikuiti^{TM}$ Enhanced Specular Reflector Film (ESR), 3M.com/displayfilms

\bibitem{sensl}Sensl C-Series SiPM, sensl.com/products/c-series

\bibitem{tpbemi}Segreto E 2015 {\it Phys. Rev.} C Evidence of delayed light emission of tetraphenyl-butadiene excited by liquid-argon scintillation light {\bf 91} 035503
\end{thebibliography}
\end{document}